\documentclass[a4paper]{aa}
\usepackage{graphicx}
\newcommand{\be} {\begin{equation}}
\newcommand{\ee} {\end{equation}}

\newcommand{\f} {\frac}
\newcommand{\pa} {\partial}

\renewcommand{\vec}[1]{\mbox{\boldmath$#1$}}
\begin{document}
\title{A numerical MHD model for the solar tachocline with meridional flow}
\author{ A. Sule \and R. Arlt \and G. R\"{u}diger}
\institute{Astrophysikalisches Institut Potsdam, An der Sternwarte 16, D-14482, Potsdam,
 Germany}
\offprints{gruediger@aip.de}
\date{Received date / Accepted date}
\authorrunning{Sule et al.}
\titlerunning{A numerical MHD model for the solar tachocline}
\abstract{
There are successful approaches to explain the formation of the tachocline by a poloidal magnetic
 field in the solar core. We present here the first MHD simulations of the solar tachocline which
 self-consistently include the meridional circulation. We show that the meridional flow significantly
 changes the shape and the characteristics of the tachocline. We find that after the inclusion of the
 meridional circulation, a tachocline can be formed even when the poloidal field lines are crossing the
 boundary between the radiative zone and the convection zone. We also discuss the effects of the
 magnetic Prandtl number as well as of the magnetic Reynolds number on the properties of the
 tachocline. The tachocline is much thinner for higher magnetic Reynolds numbers and/or lower
 magnetic Prandtl numbers. We expect that a poloidal magnetic seed field of around 1~G will be
 sufficient to produce the tachocline of the Sun. However, the model requires the initial magnetic field
 to be in a narrow range for satisfying tachocline solutions. The simulations including a stable
 temperature gradient produce a shallower as well as slower meridional circulation than the ones
 without it, as desired from the Lithium abundance at the solar surface.
}
\maketitle

\section{Introduction}
The term tachocline refers to a very thin layer at the base of the convection zone,
 which separates the latitudinal differential rotation inside the convection zone from the uniformly
 rotating radiative zone. Latest observations put the position of its center at around
 $0.69R_{\sun}$ and estimate its thickness to be slightly less than $0.05R_{\sun}$. There is
 also general belief that the tachocline is prolate and is also thicker near the pole. However,
 the observations for higher latitudes have large error bars, and hence the evidence for the
 latitudinal dependence of the tachocline is not yet conclusive (Schou et al. 1998, Antia et al.
 1998, Charbonneau et al. 1999, Basu \& Antia 2001). The facts were verified qualitatively in
 frequency-versus-radius plots presented by e.g.\ Eff-Darwich et al. (2002) and Howe (2003).

There have been many attempts to explain the formation of such a thin shear layer inside
 the Sun. \cite{SZ92} were the first to give a purely hydrodynamic model for the
 tachocline. \cite{E97} extended this analysis numerically also including meridional
 circulation. These models were successful in explaining the stability of the tachocline using
 anisotropic turbulence but the equilibrium thickness of the tachocline obtained was much thicker than
 latest observational estimates. Also later works such as that of \cite{M03} confirm that a
 purely hydrodynamic model is insufficient to explain the tachocline.

R\"{u}diger \& Kitchatinov (1997) studied the tachocline
 under the influence of a magnetic field. They believed that the existence of a weak seed
 field of about $0.01$ to 1~mG in the solar core is highly likely. And this weak internal
 magnetic field will be mainly responsible for the formation of the tachocline. Several other
 authors like \cite{GM98, MC99}, and \cite{G02} have used this idea
 of a weak poloidal seed field forming the tachocline.

Here we present a magnetohydrodynamic model, which
 self-consistently calculates the meridional flow. In this work, we show how the inclusion of the
 meridional circulation changes the shape of the tachocline and discuss the effects of varying
 magnetic Prandtl number and magnetic Reynolds number  on the structure and the properties
 of the tachocline. Finally, we discuss the effect of the stable temperature
 gradient on the structure of the meridional flow.

For a very simple magnetic shear flow model, the relation
\begin{equation}
\frac{\delta R}{R}\propto {\rm Ha}^{-1/2}
\label{delr}
\end{equation}
has been derived by R\"udiger \& Kitchatinov (1997) for the fractional thickness of any
 magnetically induced tachocline layer with the Hartmann number
\begin{equation}
{\rm Ha}=\frac{B_0 R}{\sqrt{\mu_0 \rho \nu \eta}},
\label{Ha}
\end{equation}
implying that rather thin tachoclines of say $0.03R_{\sun}$ always require ${\rm Ha}\simeq
10^3$. The Hartmann number also appears in the relation between the induced toroidal
magnetic field $B_\phi$ and the given poloidal field amplitude $B_0$, i.e.
\begin{equation}
\frac{B_\phi}{B_0} \simeq \frac{\Omega_0 R}{B_0}\left(\frac{\mu_0\rho\nu}{\eta}\right)^{0.5}
 \simeq \frac{{\rm Rm}}{{\rm Ha}}
\label{bb}
\end{equation}
(the amplification factor), with the global magnetic Reynolds number
${\rm Rm}=\Omega_0 R^2/\eta$.

With $\eta\simeq 10^3$~cm$^2$/s as a rough estimate for the solar diffusivity,
one finds ${\rm Rm}\simeq 10^{12}$ for the solar
transition region between the convection zone and the radiative interior so that a very
strong amplification of nine orders of magnitude should occur. The code
used below only reaches Rm-values of order $10^5$, thus being limited on
the amplification factors.

Forg\'{a}cs-Dajka \& Petrovay (2002) and \cite{P03} suggested that the tachocline, being just
 below the base of the convection zone, is more likely to be turbulent and hence the values of
 the diffusivities in this region are more likely to be as high as those in the convection zone.
 Studies by several authors concentrated on the stability of the tachocline being subject
 to latitudinal shear (Watson 1981, Gilman \& Fox 1997, Garaud 2001, Cally 2003,
 Dikpati et al. 2003). These ideas, however, are beyond the scope of the present discussion.

\section{The model}
The principle aim of this work will be to reproduce the observed rotation profile in the
 tachocline region, along with the nearly uniform rotation in the solar radiative zone below.

In order to be consistent with the observations, we require that (i) the thickness of the tachocline
 should be less than $0.05R_{\sun}$, (ii) the core should settle near uniform rotation, (iii)
 the meridional circulation should be slow and/or should not penetrate the radiative zone
 deeply, and (iv) the internal seed field should be weak.

We assume that the tachocline lies completely inside the radiative zone. The equations
 take axisymmetric, incompressible Boussinesq form. Though we would like to
 include density stratification for more realistic simulations, our test runs with even moderate density
 gradients proved computationally too expensive. We take
 the outer radius of the computational domain as $R_{\rm out}=0.75R_{\sun}$ and the inner radius
 as $R_{\rm in}= 0.1R_{\sun}$. Due to numerical constraints, we restrict ourselves to
 spherical shells instead of a complete sphere. The size of the remaining inner hole was found
 to have negligible effect on the results.

We also assume that the rotation profile in the convection zone is
 independent of the dynamics in the radiative zone and can be prescribed as a boundary
 condition. We further assume that any physical phenomenon occurring  in the convection
 zone, including a dynamo, does not bear an effect on the dynamics within the radiative zone
 and the tachocline. For the first set of simulations, the effect of the temperature gradient and hence
 buoyancy force on the fluid is also neglected. We will present the equations and results
 involving the temperature in Section~5.

The magnetohydrodynamic equations employed are
\begin{equation}
\f{\pa \vec{u}}{\pa t} = - (\vec{u} \cdot \nabla) \vec{u} -\nabla P + \nu \Delta\vec{u} +
\f{1}{{\mu}_{0}\rho} (\nabla \times \vec{B}) \times \vec{B},
\label{ns1}
\end{equation}
\begin{equation}
\f{\pa B_{\phi}}{\pa t} = [\nabla \times (\vec{u} \times \vec{B})]_{\phi} - \eta [\nabla
 \times (\nabla \times \vec{B})]_{\phi}
\label{ns2}
\end{equation}
with $ \nabla \cdot \vec{u}=0$ and $
\nabla \cdot \vec{B} = 0$ where $\nu$ and $\eta$ are the constant viscosity and magnetic
 diffusivity, respectively. Other symbols have their usual meaning. The
 poloidal magnetic field is maintained time-invariant. This approximation is valid as in the Sun,
 due to the small $\eta$, the magnetic diffusion time is believed to be several times
 larger than the present solar age, whereas the stable tachocline solution is reached in our
 simulations in much shorter times.

The equations are solved in a spherical shell geometry using the spectral code
 developed by \cite{H94} and \cite{H00}. The equations are scaled in time by the magnetic
 diffusion time, $\tau_{\rm diff} = R_{\rm out}^{2} / \eta$, in velocity by $\eta / R_{\rm out}$,
 and the magnetic field is normalized by $B_0$, thus generating the dimensionless Lundquist number
 \begin{equation}
 {\rm S}_0=\frac{B_0 R_{\rm out}}{\sqrt{\mu_0 \rho}\ \eta}.
 \label{S0}
 \end{equation}
 This reduces the equations to very few free parameters, namely $r_{\rm out} = R_{\rm out} /
 R_{\sun}$, $r_{\rm in} = R_{\rm in} / R_{\sun}$, Rm, $ {\rm Pm} = \nu / \eta$ and
 ${\rm S}_{0}$.

Given the very high turbulent viscosity in the convection zone, we can safely assume that
 the convection zone rotation profile acts like a rigid outer boundary on the tachocline.
 Therefore, the rotation profile of the angular velocity at the outer boundary is chosen to
 match the observed rotation profile in the bulk of the convection zone, which is also used
 by MacGregor \& Charbonneau (1999) and is maintained as a rigid boundary condition. It is
 given by
\begin{equation}
\Omega_{\rm out}= \Omega_{0} \left(1 - 0.1264 {\cos}^{2} \theta - 0.1591 {\cos}^{4}
 \theta\right)
\end{equation}
where $\theta$ is the co-latitude\footnote{\cite{MC99} give a positive sign for the $\cos^{4}\theta$ term
 which appears to be a typo. The negative sign used here is essential for a solar-like differential
 rotation.}. The remaining boundary conditions on the velocity are maintained stress-free. For the
 magnetic field, we set vacuum boundary conditions at the inner as well as the outer boundary.

Due to computational limitations, we will use values of Rm not exceeding
${10}^{5}$. We assume that the magnetic diffusivity $\eta$ and the viscosity
 $\nu$ take their molecular values in the tachocline region, but our numerically
 constrained choice of Rm implies very high values of $\eta$ and $\nu$ in the
 simulations.

The computations evolve only the axisymmetric modes of the originally three-dimensional
 spectral code. The runs typically employ 60 radial Chebyshev ($k$)
 modes and 60 latitudinal Legendre ($l$) modes. It has been verified by additional computations
 that the results change very little at different resolutions. The Chebyshev polynomials
 are ideal to resolve the radial boundary layers very well.

The magnetic field structure used for the internal seed field is dipolar with no initial
toroidal field. i.e.
\begin{equation}
 \vec{B}= \left(\f{1}{{r}^{2}\sin \theta}\f{\pa A}{\pa \theta},-\f{1}{r \sin \theta}
\f{\pa A}{\pa r},0\right)
\label{B}
\end{equation}
with the generating function
\begin{equation}
 A={\rm S}_{0} r^{2} \left(1-\f{r}{r_{\rm out}} \right) \left(1-\f{r_{\rm in}}{r} \right) \sin^{2} \theta
\label{A}
\end{equation}
The function involves the second bracket in order to avoid the magnetic field lines crossing the
 inner boundary. This will introduce small curls of the poloidal magnetic field near the inner
 boundary. We eliminate the curl of the poloidal magnetic field, which is dominant only at the inner
 boundary, by equating it to zero at every time step. A comparison with computations
 including the full Lorentz force did not show significantly different results though.


\section{The effect of the meridional flow}
Earlier investigations have shown that even a weak poloidal seed field is able to produce a solar-like
 tachocline. It was found that the region near the rotation axis (i.e. poles) is least
 affected by the field and the tachocline is thickest in that part. Further, when the poloidal field amplitude
 is very small, the magnetic field is unable to alter the rotation, and the resulting rotation
 profile looks very similar to the non-magnetic case. At the other extreme, if the magnetic field
 strength is very high, the contour lines of constant $\Omega$ follow the poloidal magnetic field
 lines throughout the interior, in accordance with the theorem of Ferraro (1937).
 A solar-like tachocline is thus impossible (Garaud 2002) in either case. Our first
 simulations neglecting the meridional flow concur with most of these key results.
\begin{figure}
\resizebox{\hsize}{!}{\hbox{\includegraphics{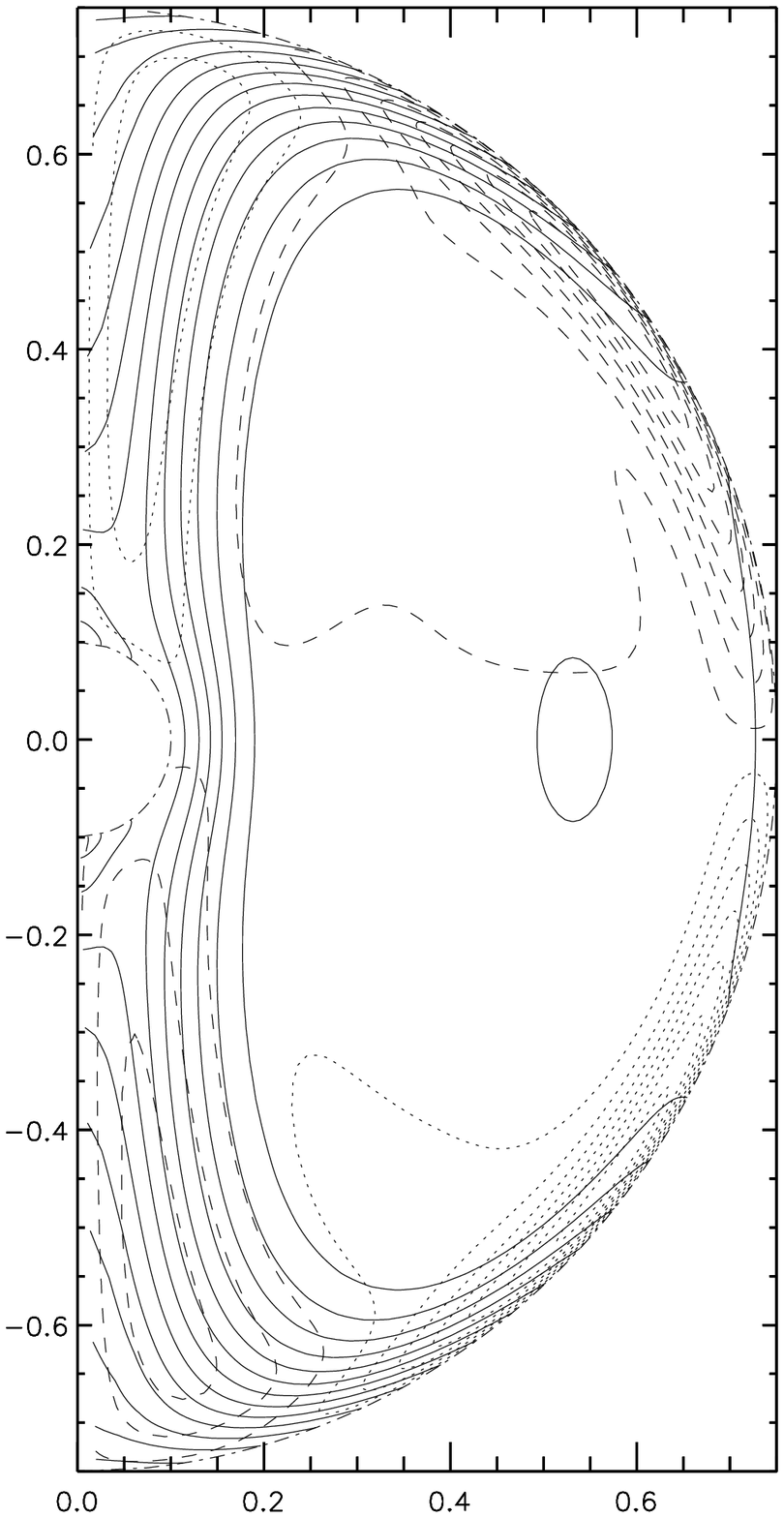} \includegraphics{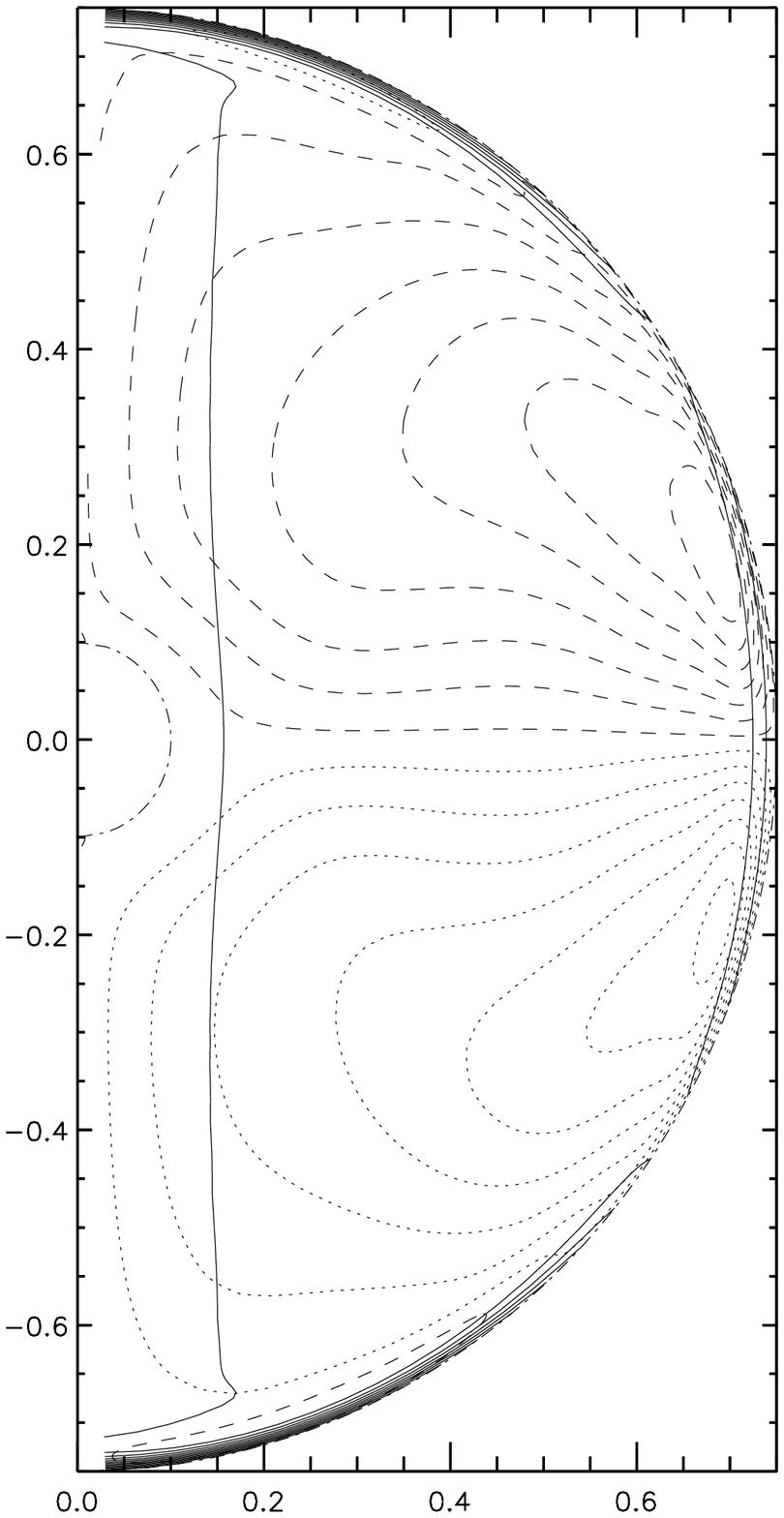}}}
\caption{Results for the simulations excluding (left) and including (right) the meridional
 circulation with the magnetic field being confined to the core. The solid lines are the
 iso-rotation curves, whereas the dashed and the dotted lines represent the contours of the
 positive and the negative toroidal field strength respectively. The dot-dashed lines indicate the
 boundaries of the simulation domain. The time-invariant poloidal magnetic field is not shown.
 ${\rm S}_0=1100$, ${\rm Pm}=1$ and ${\rm Rm}=10^4$.}
 \label{sps12}
\end{figure}

In the simulations of this Section, we use ${\rm Rm} = 10^{4}$ and ${\rm Pm}=1$. In all
 our simulations, we found that the system converges to an equilibrium solution in less than
 $0.1\tau_{\rm diff}$. Unless mentioned otherwise, the figures and numerical values throughout
 this Paper refer to the final equilibrium solutions. The results are best represented by the
 comparison of the following two graphs. While the left panel of Fig.~\ref{sps12} is just a reproduction
 of the earlier MacGregor \& Charbonneau (1999) result, where the magnetic fields are confined to
 the simulation domain, the right panel of Fig.~\ref{sps12} shows the same model including
 the meridional circulation. The solid lines are the iso-rotation curves whereas the dashed
 and the dotted lines represent the contours of the positive and negative toroidal field strength,
 respectively. While in the left graph the region along the rotation axis is not affected at
 all, we see that the inclusion of the meridional circulation changes the picture completely.

We observe that the entire core, including the region near the rotation axis, has achieved nearly
 uniform rotation. The tachocline is formed near the outer boundary. In contrast with the results
 ignoring the meridional flow, the tachocline is now thinnest near the pole. In the region near
 the equator, where the magnetic field influence is smaller, we find that the iso-rotation
 curves tend to be similar to the characteristic Taylor-Proudman flow. We further observed
 that the toroidal magnetic field strength is only 30\% of the poloidal magnetic field strength.

\begin{figure}
\resizebox{\hsize}{!}{\hbox{\includegraphics{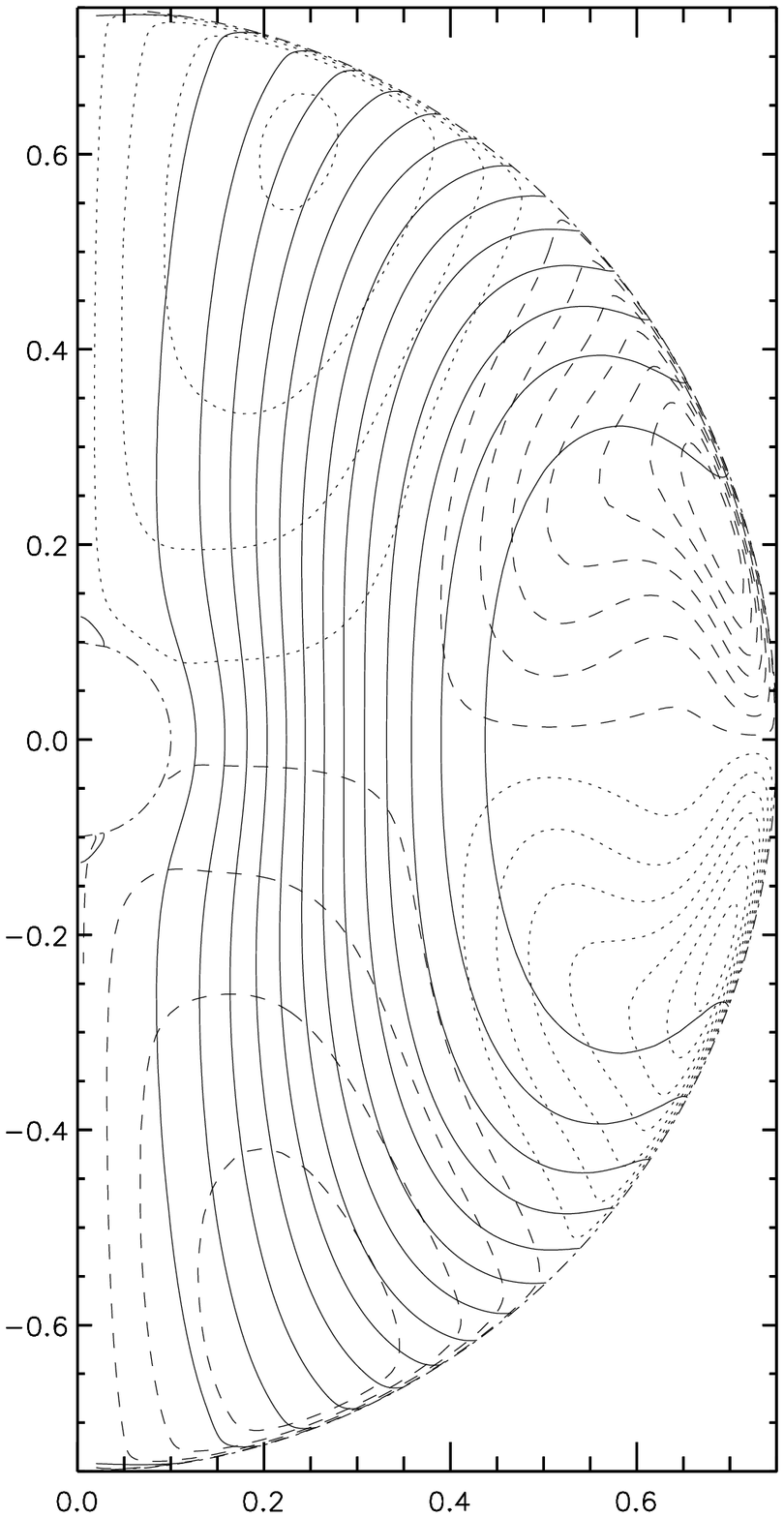} \includegraphics{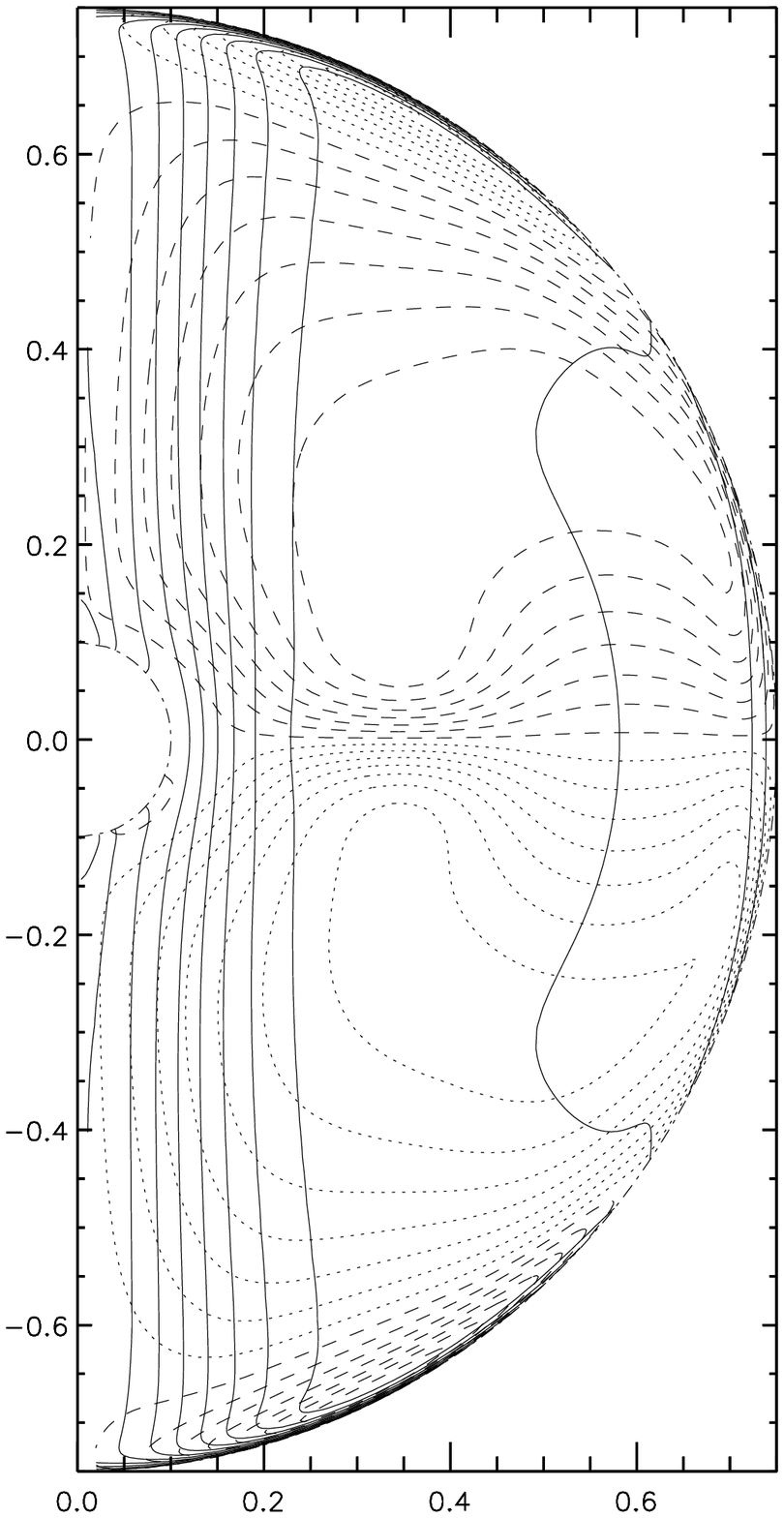}}}
\caption{Results for the simulations excluding (left) and including (right) the meridional
 circulation when the magnetic field lines are crossing the outer boundary.}
\label{sps34}
\end{figure}

\begin{figure}
\resizebox{\hsize}{!}{\includegraphics{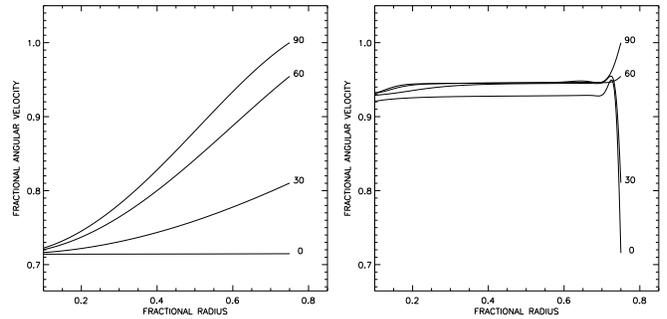}}
\caption{Fractional $\Omega$ vs. fractional radius for the initial configuration (left)
 and the equilibrium solution when $B_{\rm pol}$ is constant (right). Different
 lines represent $\Omega$ at different co-latitudes.}
\label{dmagcomp}
\end{figure}

Following MacGregor \& Charbonneau (1999), we also studied the second case, where the
 poloidal magnetic field is no more confined to the computational domain and the magnetic field
 lines are crossing the outer boundary of the simulation domain. While this case produced no real
 tachocline for the MacGregor \& Charbonneau (1999) model, this is no longer a problem after
 the inclusion of the meridional flow (see Fig.~\ref{sps34}). The initial and final states
 in terms of fractional $\Omega$ are shown as functions over radius for various latitudes
 in Fig.~\ref{dmagcomp}.

\begin{figure}
\centerline{\resizebox{\hsize}{!}{\includegraphics{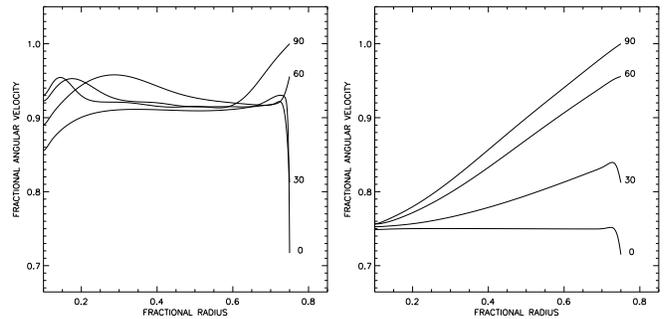}}}
\caption{Fractional $\Omega$ vs. fractional radius for a snapshot at $0.05\tau_{\rm diff}$ (left)
 and $0.5\tau_{\rm diff}$ (right) with decaying $B_{\rm pol}$. Again, different lines represent
 $\Omega$ at different co-latitudes.}
\label{dmagcomp_decay}
\end{figure}

\begin{figure}
\resizebox{\hsize}{!}{\hbox{\includegraphics{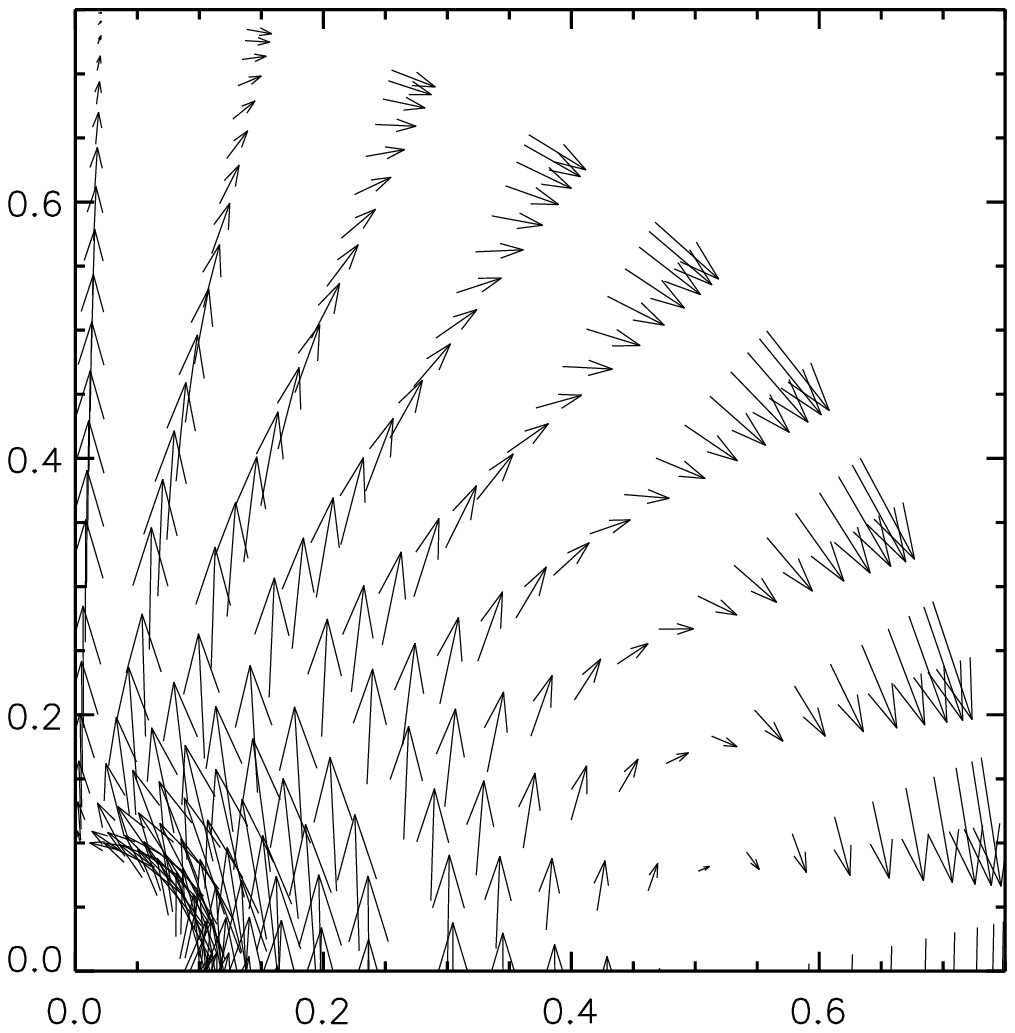} \includegraphics{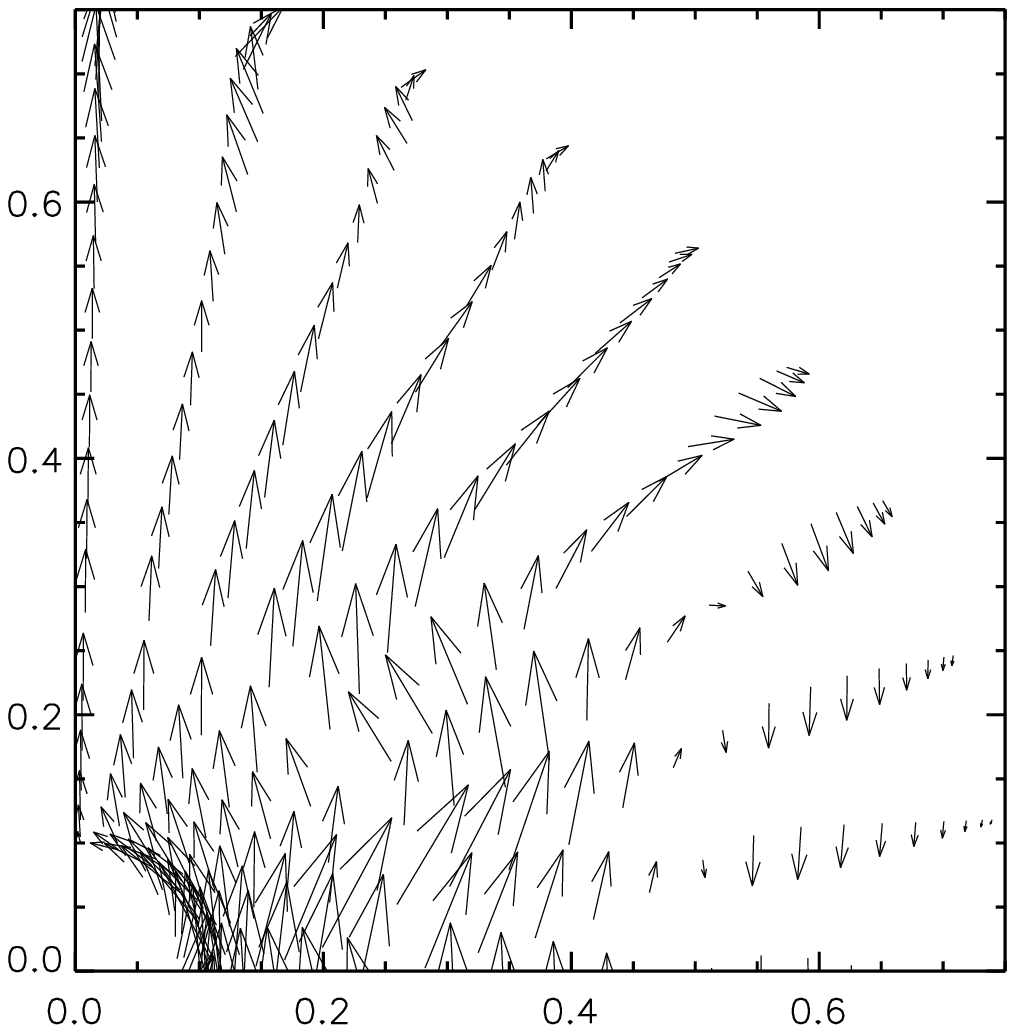} \includegraphics{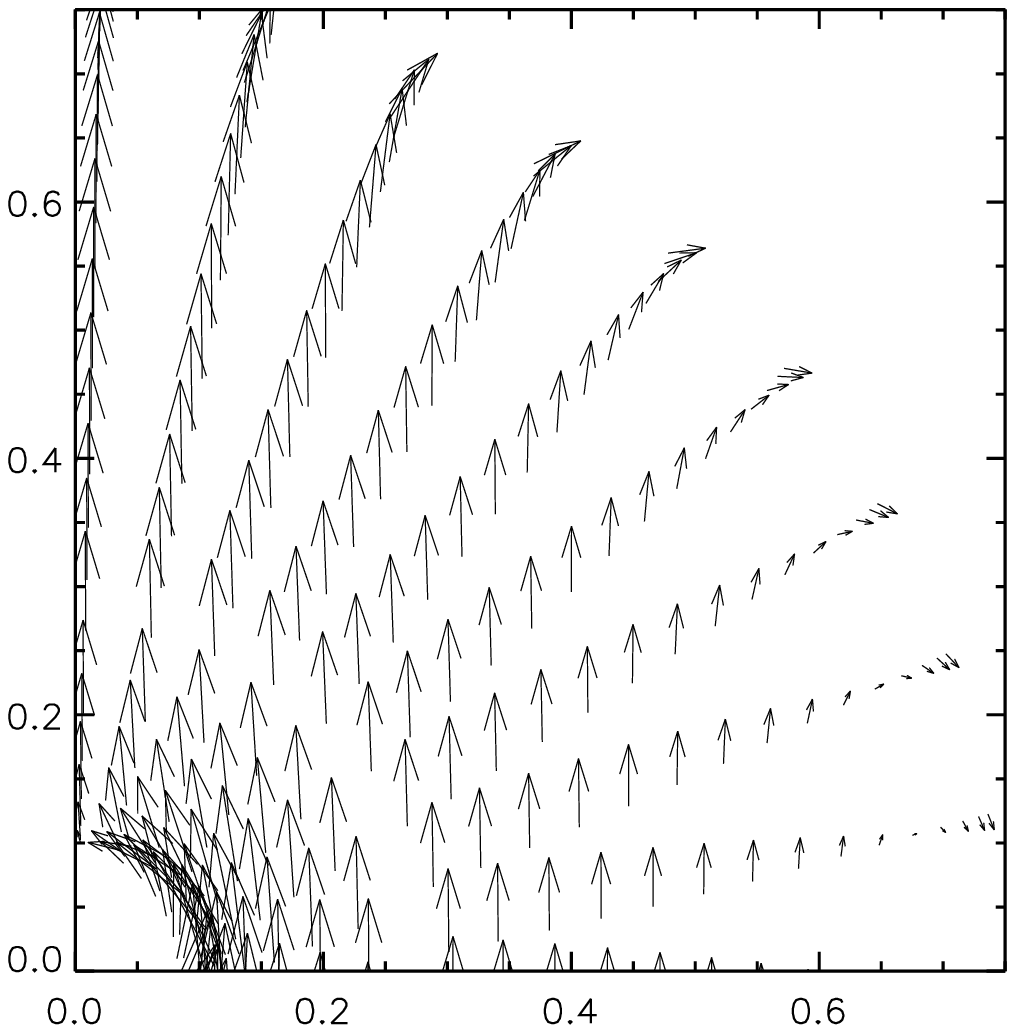}}}
\caption{Snapshots of $B_{\rm pol}$ in case of a decaying poloidal field at (from left to right)
 $t = 0$, $0.05\tau_{\rm diff}$, and $0.5\tau_{\rm diff}$. For every snapshot, length of arrows is
 proportional to the amplitude of $B_{\rm pol}$ at each point. The length of arrows is renormalized
 in each plot by corresponding maximum field amplitude. Only every fifth arrow in $\theta$ and $r$
 is plotted in order to avoid congestion.}
\label{dmagsps}
\end{figure}

We performed some test simulations with a decaying poloidal magnetic field. During these
 simulations, we observed that the magnetic field does form a tachocline during the period up to
 $\sim0.05\tau_{\rm diff}$, but as the magnetic field decays gradually due to the high value of
 $\eta$ assumed in the simulations, the flows readjust themselves, and the resulting pattern will be the
 same as the purely hydrodynamic case, i.e. it will approach the Taylor-Proudman flow.
 Fig.~\ref{dmagcomp_decay} shows the fractional $\Omega$ in an intermediate stage at
 $t=0.05\tau_{\rm diff}$ for a decaying poloidal magnetic field, keeping other parameters the same
 as for the run in Fig.~\ref{dmagcomp}. The effect of forming a tachocline structure by a
 time-dependent poloidal field is thus very similar to the non-decaying field, but diffusivity reduces the
 field unnaturally early, and the final tachocline state is not achieved. Fig.~\ref{dmagsps} shows the
 evolution of the poloidal magnetic field at three different times.

We conclude here that the magnetic field is not only important for the formation of the
 tachocline but is also important for maintaining it, at least in the case of the high magnetic
 diffusivity used in these simulations.

\section{Effect of varying magnetic Prandtl number and magnetic Reynolds number}

\subsection{Varying the magnetic Prandtl number}

Following Kippenhahn \& Weigert (1994) and Stix \& Skaley (1990), the solar value for the
 magnetic Prandtl number is of the order of
 $10^{-3}$. Although this value was not achieved, the simulations were performed for
 various values of Pm in the range of 0.05--1 whereas ${\rm Rm}=10^{4}$ in all the
 cases.

\begin{figure}
\resizebox{\hsize}{!}{\hbox{\includegraphics{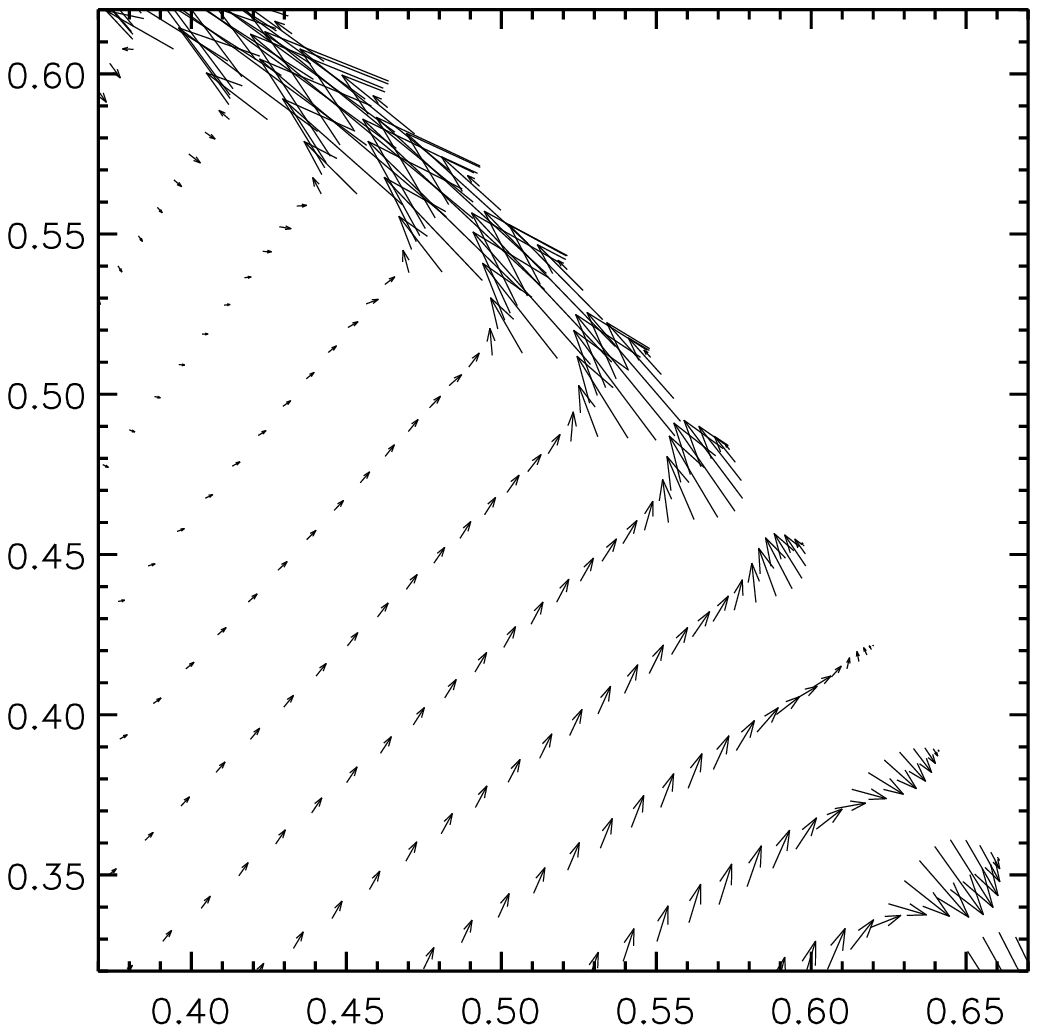} \includegraphics{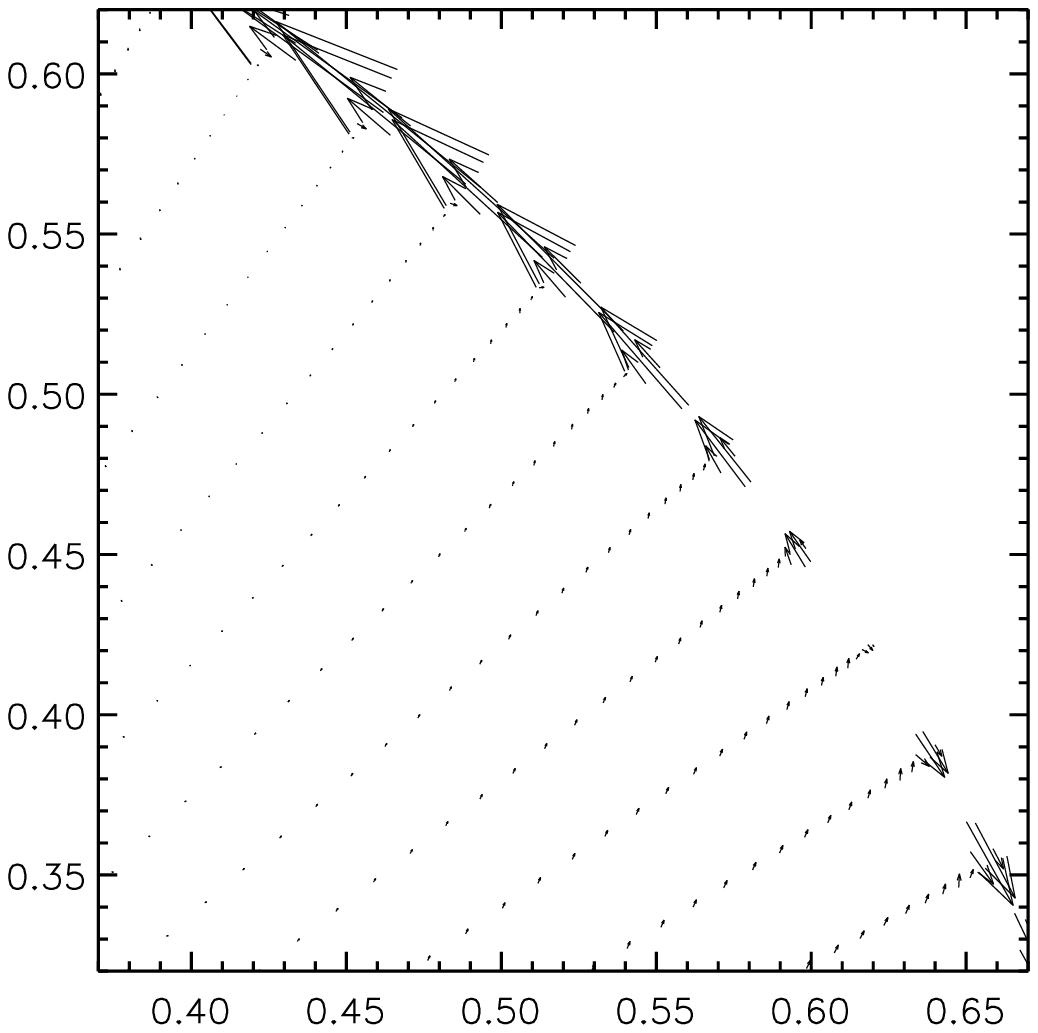}}}
\caption{Results for the simulations with ${\rm Pm}=1$ (left) and 0.05 (right) showing a
 small section of the simulation domain with the components of the meridional circulation
 plotted. The arrow lengths are assigned in the same way as Fig.~\ref{dmagsps}. Only alternate
 points in the radial direction are plotted.}
 \label{sps56}
 \end{figure}

In these simulations, we observed that the horizontal component of the meridional flow
 ($u_{\theta}$) generated is about 2\% of $u_{\phi}$ at higher latitudes and is about 1\% for
 lower latitudes. Similarly, the vertical component of the meridional flow ($u_r$) is about
 0.5\% (see Fig.~\ref{uratio3} left). These ratios are nearly independent of the value of Pm.
 The radial component of the velocity remains much smaller than the latitudinal
 component especially for lower values of Pm, as can be seen in Fig.~\ref{sps56}. The Figure also
 shows that strong latitudinal flows are expelled closer to the outer boundary for
 lower values of Pm. The tachocline, thus formed, is thinner\footnotemark \  at the pole
 than at the equator. The thickness of the tachocline at the equator reduces marginally from
 $0.030R_{\sun}$ to $0.026R_{\sun}$ when Pm is changed from 1 to 0.05, whereas the
 thickness of the tachocline at the pole decreases considerably. We also observe that the
 amplitude of the toroidal magnetic field remains smaller than that of the poloidal field and is
 almost independent of the choice of Pm.
 \footnotetext{The thickness is defined as the distance between the outer boundary and the
 radius at which the rotation rate deviates by 1\% from the rotation rate deep inside the
 core.}

\subsection{Varying the magnetic Reynolds number}

\begin{figure}
\resizebox{\hsize}{!}{\includegraphics{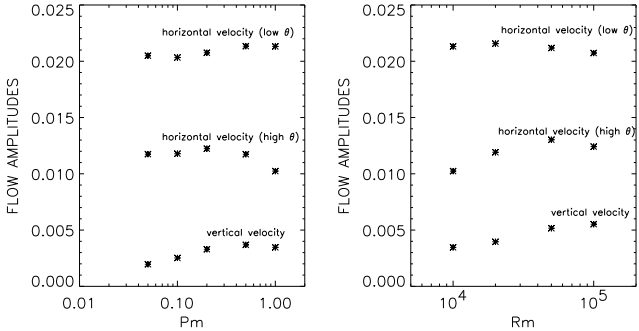}}
\caption{The amplitudes of the components of the meridional flow at different
 co-latitudes ($\theta$), normalized to the rotational velocity at the equator, with varying Pm
 (left) and Rm (right).}
\label{uratio3}
\end{figure}

Now we will use a constant ${\rm Pm}=1$ and vary Rm. As we know that
 the solar value of Rm is about $10^{12}$, we ran simulations for
 Rm larger than ${\rm Rm}=10^{4}$. The highest value of Rm in our
 simulations was $10^{5}$.

In this set of simulations we again notice that the relative amplitudes of the meridional flow are nearly
 independent of the value of Rm as shown in Fig.~\ref{uratio3} (right). The thickness of the
 tachocline at the equator reduces from $0.030R_{\sun}$ to $0.015R_{\sun}$ when Rm is
 changed from $10^{4}$ to $10^{5}$, whereas the thickness of the tachocline at the pole goes
 down from $0.012R_{\sun}$ to $0.003R_{\sun}$. Correspondingly, the toroidal field
 ($B_{\phi}$) in the tachocline region increases from $0.3 B_0$ to $1.3 B_0$.
\subsection{Effect on the Lundquist number}
\begin{figure}
\centerline{\resizebox{0.8\hsize}{!}{\includegraphics{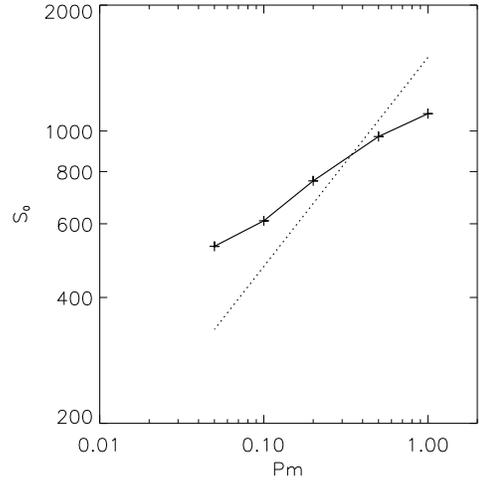}}}
\caption{Variation of ${\rm S}_{0}$ with varying Pm as a log-log plot, for an approximate
 tachocline thickness of $0.025R_{\sun}$. The dotted line represents a fixed ${\rm Ha}=1500$.}
\label{uratio5}
\end{figure}

\begin{figure}
\centerline{\resizebox{0.8\hsize}{!}{\includegraphics{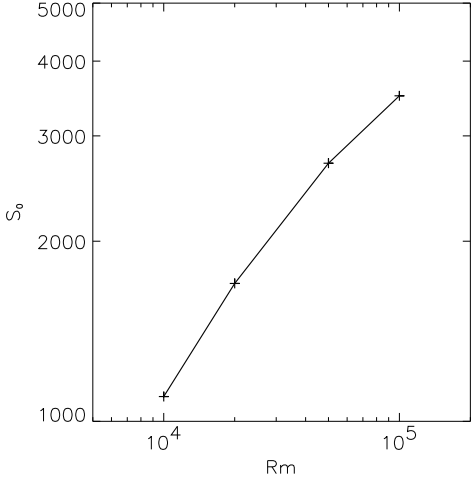}}}
\caption{Variation of ${\rm S}_{0}$ with varying Rm as a log-log plot, for an approximate
 tachocline thickness of $0.028R_{\sun}$.}
\label{uratio4}
\end{figure}

It will be worthwhile to give some emphasis on the choice of the poloidal field strength.
 In our simulations, we find that the choice of the amplitude of the seed field value is very critical.
 A small deviation in either direction from the magnetic field which produces a solar-like tachocline
 either makes the iso-rotation curves similar to the non-magnetic
 case or the curves will comply with Ferraro's theorem, unable to produce a tachocline
 in both cases. At lower values of Pm, we require smaller seed field as
 shown in Fig.~\ref{uratio5}. At higher values of Rm, the required value of
 ${\rm S}_{0}$ is higher (Fig.~\ref{uratio4}).

An empirical law governing the variation of $S_{0}$ as a combined function of Pm and Rm can be
 deduced from these plots as
\begin{equation}
{\rm S}_{0} \simeq 10\  {\rm Rm}^{0.5}{\rm Pm}^{0.25}.
\label{scal}
\end{equation}
This scaling is the main result of the computations. It means that
\begin{equation}
\frac{B_0}{\sqrt{\mu_0\rho}}\simeq 10  \root 4 \of{\Omega^2\nu \eta}
\label{bmurho}
\end{equation}
which, of course, has the correct dimension of a velocity. The old estimate without meridional
 flow R\"udiger \& Kitchatinov (1997) led to the quite different expression $B_0/\sqrt{\mu_0
 \rho}\simeq 10^3 \sqrt{\nu \eta}/R$ which is a very small value due to the appearance of the
 radius $R$. Eqn.~(\ref{bmurho}) yields
\begin{equation}
\frac{B_0}{\sqrt{\mu_0 \rho}}\simeq 0.1~{\rm cm/s},
\label{bnull}
\end{equation}
so that a maximum field amplitude of 1~Gauss results, for an average density of 10~g/cm$^3$.
 This is a much larger value than the mGauss values for models without meridional flow, but
 it is not an unrealistic number. By contrast with the old model, the toroidal field belts now
have the same order of magnitude as the poloidal fields.
\section{Effect of a temperature gradient}
In the simulations discussed in the previous sections, we noted that the amplitude of the
 meridional circulation was nearly independent of the variation of Rm as well as Pm. In the
 lower latitudinal belt, where the solar dynamo is likely to be located, the amplitude of horizontal
 velocity $u_{\theta}$ was always around  1\% of $u_{\phi}$, and the flow reached very deep layers
 of the shell. The Lithium abundance, however, as observed at the solar surface suggests
 that the meridional circulation should either be very shallow or very slow. Otherwise Lithium
 would be destroyed in its fusion zone below $0.68R_{\sun}$.

We therefore include a given temperature gradient in our model. We introduce a
 temperature fluctuation $\Theta$ on top of this temperature profile. This fluctuation
 induces a buoyancy force. Hence Eqn.~(\ref{ns1}) will be modified to read
 \begin{eqnarray}
 \lefteqn{ \f{\pa \vec{u}}{\pa t} = -(\vec{u} \cdot \nabla) \vec{u} -\nabla P + \nu {\Delta}
 \vec{u} + \f{1}{{\mu}_{0}\rho} (\nabla \times \vec{B}) \times \vec{B}} \nonumber \\
 && \quad\quad + \vec g\rho, \label{t2}
\end{eqnarray}
where $\vec g$ is the vector of gravitational acceleration. In the context of the Boussinesq
approximation and after rescaling as in Section~2, we obtain a buoyancy
force as ${\rm \tilde{R}a}\Theta \vec{r}$ with the modified Rayleigh number
 \begin{equation}
{\rm \tilde{R}a}=\frac{g \alpha \Delta T R_{\rm out}^3}{\eta^2},
\end{equation}
where $\Delta T = T_{\rm in}-T_{\rm out}$ and $\alpha$ is the coefficient of volume expansion.
The equilibrium solution obtained from the temperature equation alone is chosen as
 the mean temperature profile (normalized with its bottom value) and is given by
\begin{equation}
T_0= \frac{r_{\rm in}}{r_{\rm out}-r_{\rm in}} \left(\frac{r_{\rm out}}{r}-1\right)
\label{T0}
\end{equation}
 and the non-dimensional energy equation has the form
\begin{equation}
\frac{\partial \Theta}{\partial t} = \frac{{\rm Pm}}{{\rm Pr}} \Delta \Theta - \vec{u} \cdot \nabla (\Theta+T_0),
\label{partheta}
\end{equation}
where ${\rm Pr}= \nu/\chi$ is the Prandtl number using the thermal diffusivity $\chi$.
\begin{figure}
\resizebox{\hsize}{!}{\hbox{\includegraphics{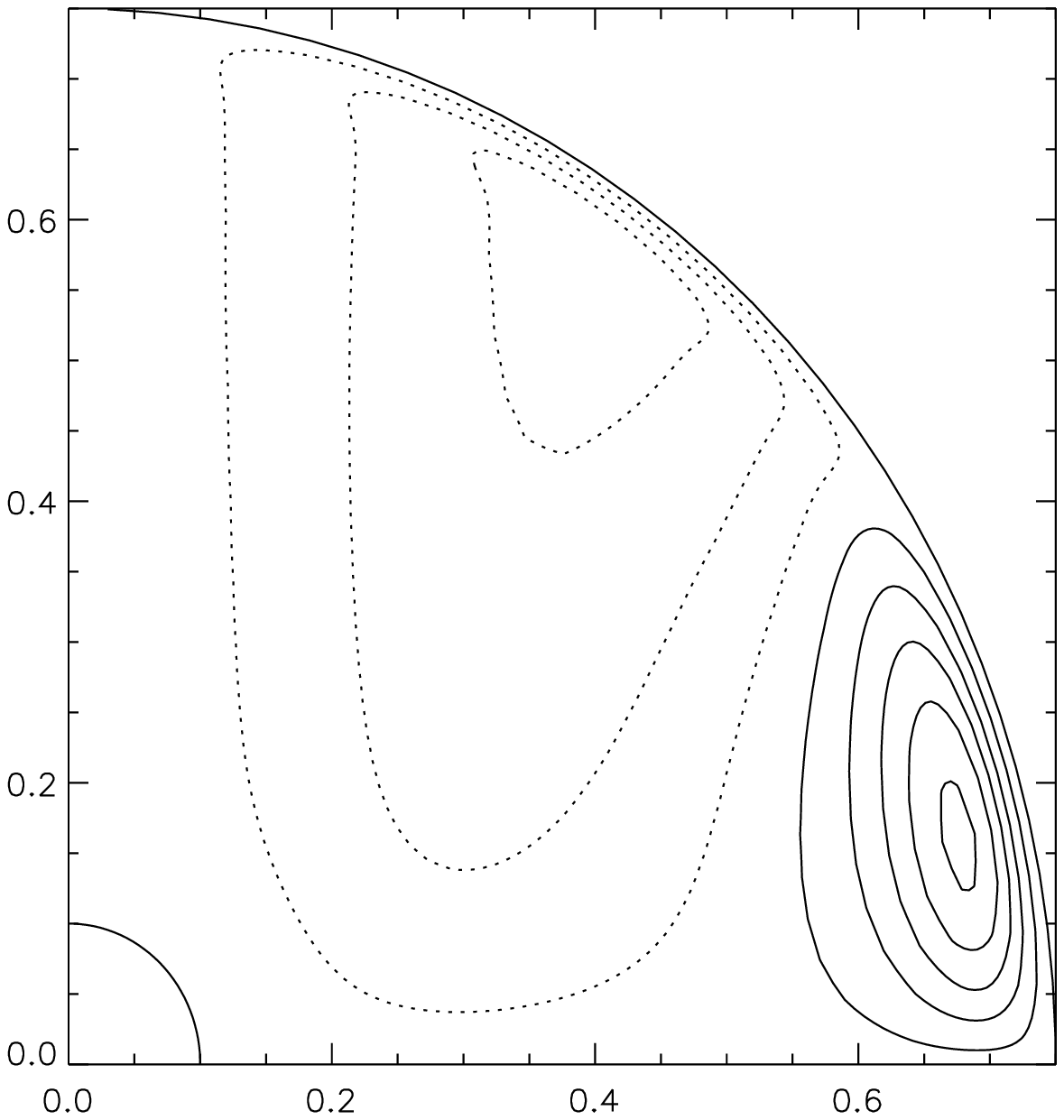} \includegraphics{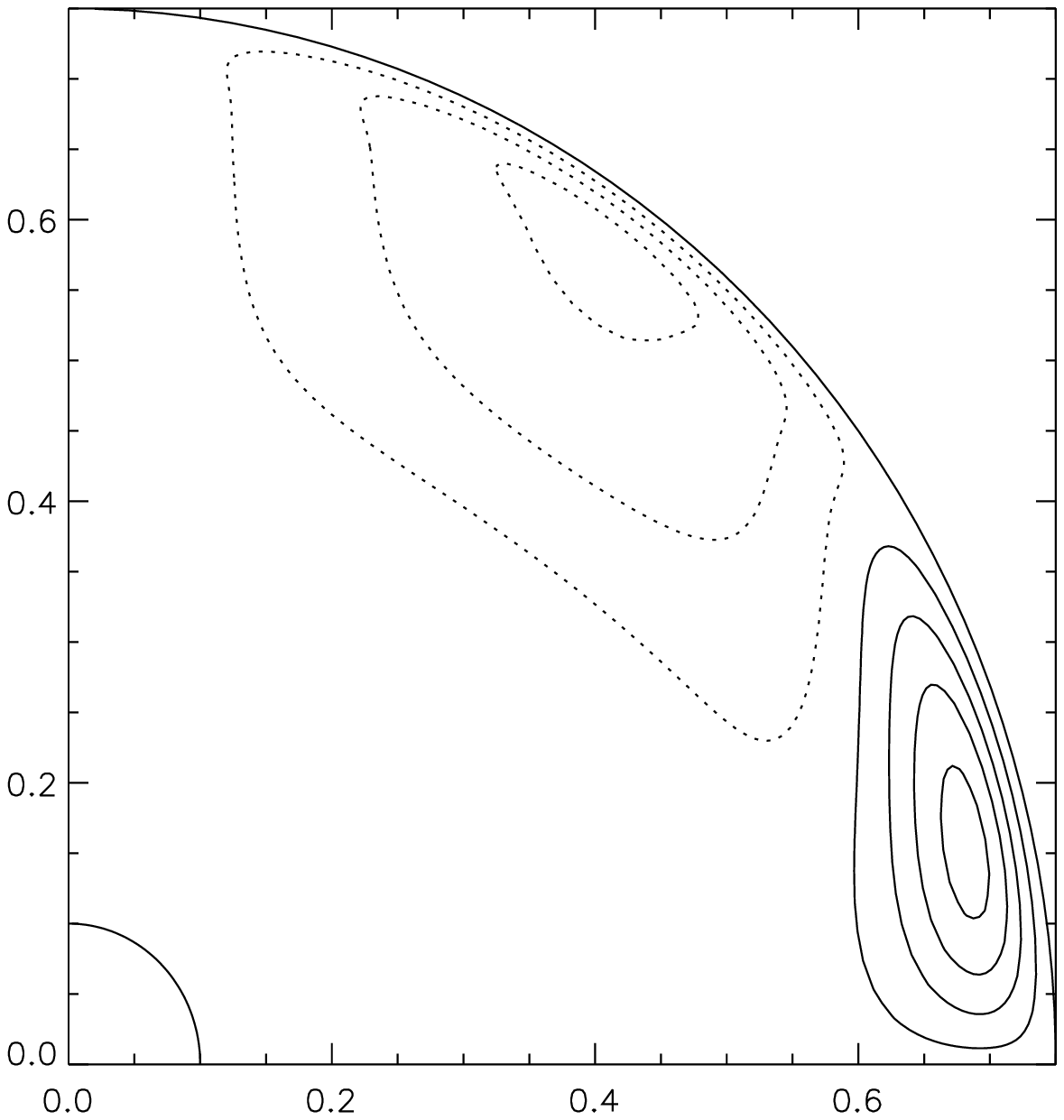}}}
\caption{Results for the simulations excluding (left) and including (right) buoyancy.
 The solid and the dotted lines are clockwise and anti-clockwise meridional
 circulation, respectively. ${\rm \tilde{R}a}= -2 \times 10^{7}$, ${\rm Rm}=10^4$.}
 \label{pot}
\end{figure}

When the model is evolved from the initial state including the temperature equation, it takes much
 longer (in terms of magnetic diffusion times) to achieve a steady state solution than the simpler cases
 in the previous sections. To save the computational resources, we fed the steady state
 solutions obtained in the previous sections as the initial condition for the simulations
 involving temperatures. It was verified that the solutions obtained in this manner are identical to
 the solutions obtained from same buoyancy runs but starting with the very initial conditions of
 Section~2.

\begin{figure}
\begin{center}
\resizebox{0.5\hsize}{!}{\hbox{\includegraphics{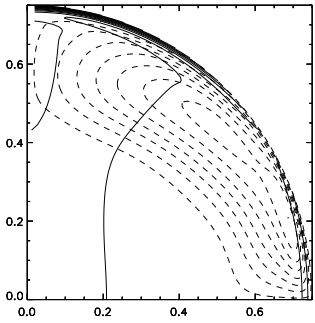}}}
\end{center}
\caption{The same as in Fig. \ref{sps12} but with the effect of buoyancy, ${\rm Ra}=-2\times10^7$,
 ${\rm Rm}=10^4$.}
 \label{sps13}
\end{figure}

\begin{figure}
\resizebox{\hsize}{!}{\includegraphics{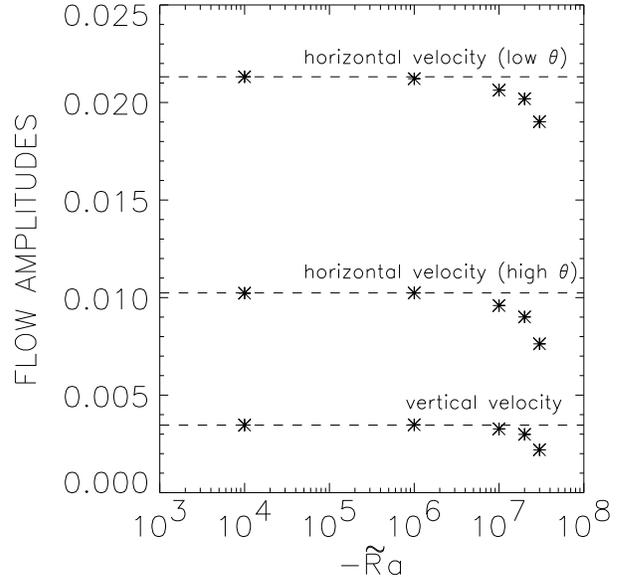}}
\hspace{-3.0 cm}
\caption{Plot of the amplitudes of the components of the meridional flow at different co-latitudes
 $\theta$, normalized to the rotational velocity at the equator, with varying ${\rm \tilde{R}a}$.
 The dashed lines represent the corresponding values without buoyancy effect.
 ${\rm Rm}=10^4$, ${\rm Pm} =1$.}
\label{uratio6}
\end{figure}

We used ${\rm Rm}=10^{4}$ and ${\rm Pm}=1$ for these sets of simulations. As the
 simulation domain is of radiative nature, we use negative values of ${\rm \tilde{R}a}$. We
 performed the simulations with various values of ${\rm \tilde{R}a}$. The results are presented
 in Fig.~\ref{pot}. For the regime when $-{\rm \tilde{R}a}/{\rm Rm}^{2} \leq 0.01$, the buoyancy
 is unable to produce any major change in any of the velocity components in magnitude or in
 structure. For more negative ${\rm \tilde{R}a}$, we see a gradual change in the structure of the
 meridional circulation\footnote{Note that ${\rm -\tilde{R}a}/{\rm Rm}^2\simeq g/ (\Omega^2 R)$}.
 The circulation is then much shallower for higher latitudes (low $\theta$), as desired to explain weak
 mixing into the interior. The magnitude of the flow also decreases but the change is not drastic; see
 Fig.~\ref{uratio6}. For the lower latitudinal belt (high $\theta$), which is an important region
 for the solar dynamo, the decrease is stronger in a relative way and the depth of the circulation
 is also clearly reduced. We get a marginal increase in the amplification of the toroidal field,
 probably because the field is not advected through the entire computational domain anymore.
 The structure of the toroidal field is thus changed as well, and it is shifted towards the outer
 parts of the shell. On the other hand, the rotation rate at higher latitudes, at large depth in
 the core, becomes slightly slower than that at the equator. But even this change is marginal and
 within the observational limits ($\vert1 - \Omega_{\rm pole} / \Omega_{\rm equator}\vert \leq 3$ \% for
 $r \leq 0.65R_{\sun}$). Angular velocity and toroidal field belts are shown in Fig.~\ref{sps13}.
 We expect that even more negative ${\rm \tilde{R}a}$ will further reduce the depth and amplitude
 of the meridional circulation.

\section{Summary}

In this work, we have presented the first MHD model for the solar tachocline which
 self-consistently calculates the meridional circulation. The consideration of the
 meridional flow changes the shape, the structure and the characteristics of the
 tachocline radically. Hence, the meridional circulation cannot be neglected while modeling
 the solar tachocline. The thickness of the tachocline in the outer boundary layer near the
 equator will be determined by the gradient of the magnetic field near the
 boundary. Our simulations show that the tachocline is thinner near the pole. But the
 simulation domain has uniform values of viscosity and magnetic diffusivity, and it will
 only be fair to say that we simulate that part of the tachocline, which is inside the radiative
 zone. Hence, a definitive conclusion about the thickness of the tachocline in the polar region
 cannot be drawn from these simulations.

We further report that the tachocline is thinner at lower values of Pm as well as at
 higher values of Rm. The meridional circulation is nearly independent of the variation of
 Rm and Pm. The amplification of the toroidal magnetic field is naturally larger at higher values
 of Rm, whereas from (\ref{bmurho}) it is clear that the poloidal magnetic field
 amplitude required to produce a solar-like tachocline goes down with decreasing $\eta$, i.e.
 increasing Rm. The magnetic seed field required to produce a solar-like tachocline is a function of
 Rm as well as Pm. The value of the seed field is expected to be around 1~Gauss in the Sun, for an
 average density of 10~g/cm$^3$. Again a
 significant change is noted from the simulations without meridional flow where even a
 sub-mGauss field was enough to produce a solar-like tachocline. The scaling of the
 magnetic field as a function of the rotation rate, $\nu$, and $\eta$ is given in (\ref{bmurho}).
 The toroidal magnetic field is expected to be a few orders higher than the poloidal
 seed field in the case of the Sun.

When a stable temperature gradient is introduced across the shell, it makes the meridional circulation
 shallower as well as weaker for stronger stabilization (more negative Rayleigh number). This can
 prevent Lithium from reaching its fusion zone which starts just below tachocline. The stabilizing effect
 is in line with the relatively high Lithium abundance observed at the surface of the Sun.
 The toroidal magnetic field in this case is limited to belt in the outer parts of the radiative zone.

The relations describing the variation of various parameters such as the Lundquist number
 required to form a solar-like tachocline, the amplification of the toroidal magnetic field, the
 amplitude of the meridional circulation etc.\ are based on the simulation results for a limited
 range of ${\rm Rm}$ and ${\rm Pm}$. We hope to improve the code and verify the correctness
 of these relations closer to the solar values of ${\rm Rm}$ and ${\rm Pm}$ in the near
 future.

\begin{acknowledgements}
L. Kitchatinov is acknowledged for the discussions about the simulation results and H.M. Antia
 for clarifying the observational results. J.-P. Zahn is acknowledged for going through 
 the manuscript and giving valuable suggestions.
\end{acknowledgements}

\end{document}